Liposomic lubricants suppress shear-stress induced inflammatory gene regulation in the joint in vivo.

Linyi Zhu[1,†,*], Weifeng Lin[2,†], Monika Kluzek[2,†], Jadwiga Miotla-Zarebska[1], Vicky Batchelor[1], Matthew Gardiner[1], Chris Chan[1], Peter Culmer[3], Anastasios Chanalaris[1], Ronit Goldberg[2], Jacob Klein[2,+,*], Tonia L. Vincent[1,+,]

[1] Kennedy Institute of Rheumatology, Centre for OA Pathogenesis Versus Arthritis, University of Oxford, Roosevelt Drive, Oxford OX3 7FY, UK

[2] Dept.of Molecular Chemistry and Materials Science, Weizmann Institute of Science, Rehovot 7610001, Israel

[3] School of Mechanical Engineering, University of Leeds, UK

† - These authors contributed equally

* linyi.zhu@kennedy.ox.ac.uk, jacob.klein@weizmann.ac.il

\+ Equal contribution

**Abstract**

Osteoarthritis (OA) is a widespread, debilitating joint disease associated with articular cartilage degradation. It is driven via mechano-inflammatory catabolic pathways, presumed up-regulated due to increased shear stress on the cartilage-embedded chondrocytes, that lead to tissue degeneration. Here we demonstrate that the up-regulation of the matrix metalloproteinase 3 (*Mmp3*) and interleukin-1β (*Il1b*) genes upon surgical joint destabilisation in a model of murine OA is completely suppressed when lipid-based lubricants are injected into the joints. At the same time, *Timp1*, a compression but not shear-stress sensitive gene, is unaffected by lubricant. Our results provide direct evidence that biolubrication couples to catabolic gene regulation in OA, shed strong light on the nature of the chondrocytes' response to shear stress, and have clear implications for novel OA treatments.

**Introduction:**

In healthy joints, the cartilage surfaces are exquisitely lubricated to allow near frictionless motion during articulation(*1-3*). Injury to the joint surface or destabilisation of the joint due to ligamentous injury leads to increased shear stress on the superficial cartilage and drives degradative processes that are associated with the development of osteoarthritis (OA) (*4, 5*),(*6, 7*). However, the nature of the stress, and of the cells within the joint that respond to it, are unknown.

Acute injurious mechanical stresses, including those induced in vivo by surgically destabilising the knee joint of mice, drive inflammatory signalling and induce inflammatory genes in chondrocytes, the cells of the articular cartilage(*8-11*), including matrix degrading enzymes that contribute to the breakdown of cartilage in OA(*12, 13*). Previous studies have shown that inflammatory gene regulation in the whole joint occurs rapidly, within 6h of joint destabilisation, but can be abrogated if the joint is completely immobilised (by general anaesthetic or by sciatic neurectomy) immediately after surgery, leading to protection against development of OA(*14*). Such immobilisation minimises shear stress by eliminating joint articulation, but allows compressive load to occur on walking. An alternative, which in principle could have a similar effect to joint immobilisation, would be to reduce shear stress by reducing the friction between the cartilage surfaces as they slide past each other. This reduction should correspondingly reduce their catabolic gene regulation(*1*). In this study we demonstrate the validity of this concept.

While friction of articular cartilage has been studied for decades(*15-17*), and a range of mechanisms suggested to account for its remarkably low value(*16, 18-22*), recent studies have increasingly emphasized the role of hydration lubrication(*23-26*) by boundary layers exposing highly-hydrated lipid-headgroups at the cartilage surfaces(*1, 2*). Indeed it has been explicitly suggested(*1*) that intra-articular (IA) injection of suitable lipid vesicles would augment cartilage lubrication (i.e. reduce the friction) and thus help to supress the catabolic gene upregulation underlying OA progression. A number of recent studies(*27, 28*) attempted to examine this idea through IA injection of lubricious nanoparticles, further incorporating anti-inflammatory drugs. In these, however, any regulation in the expression of catabolic enzymes could be traced to the presence of the incorporated drugs, demonstrating that their efficacy arises from the drugs they delivered(*27, 28*) rather than to lubrication by the nanoparticles. In the present work, in contrast, we determine the effect of lubrication alone (by liposomes) on the chondrocyte gene regulation.

We examine the lubricating properties of our functionalized liposomes on layers mimicking the outer surfaces of articular cartilage, then measure their retention times in the joint cavity following IA injection, and, separately, their localization on the cartilage surface. With respect to gene-regulation, among the genes that are known from previous studies to be regulated within articular chondrocytes upon joint destabilisation we focus on a panel of mechano-inflammatory genes with roles in cartilage turnover: matrix metalloproteinase 3 (*Mmp3*) and interleukin 1 beta (*Il1b*), as well as a gene that is mechanically induced in articular cartilage by compressive rather than shear stress, TIMP metallopeptidase inhibitor 1 (*Timp1*). We validate our findings, obtained by quantitative PCR of whole joint and microdissected articular cartilage, using an RNAscope approach which enables visualization of the relevant gene expression in individual cells. This enables us to identify exactly which cells in the cartilage are responding to shear stress and lubricant in vivo.

**Results.**

Experiments were carried out in two laboratories (the Kennedy Institute, University of Oxford, and the Weizmann Institute, Rehovot) and care was taken to use the same murine samples (10 – 11 weeks-old male C57BL/6 mice) and the same poly(2-methacryloyloxyethyl

phosphorylcholine)(pMPC)-functionalized (i.e. pMPCylated-) liposomes throughout in both labs (*29*).

Size distribution, morphology and biocompatibility of pMPCylated-liposomes

Large unilamellar vesicles (LUVs) were prepared from hydrogenated soybean phosphatidylcholine (HSPC) lipids functionalized by 2% (mol/mol) pMPC, designated HSPC-pMPC-LUVs. Dynamic light scattering (DLS) measurements (*29*) of the liposomes show a uniform size and narrow distribution with an average size around 170 nm (Fig. S1A), confirmed *via* Cryo-TEM imaging (*29*), where we observe distinct unilamellarity and vesicular morphology (Fig. S1B). The HSPC-pMPC-LUVs were checked for biocompatibility, showing no significant effect on cell viability (Fig. S2).

Lubrication by HSPC-pMPC-LUVs on cartilage mimicking surfaces

As proposed earlier(*2, 30-35*), the lubricating boundary layers on cartilage comprise lipids (in lamellar or vesicular form) complex with hyaluronic acid (HA) exposed at the outer cartilage surface. We therefore created HA-exposing surfaces by first coating freshly-cleaved, atomically-smooth bare mica with positively-charged poly(allylamine hydrochloride) (PAH), followed by incubation in HA solution. After washing away excess HA, HSPC-pMPC-LUVs were added to attach to the surface-attached HA layer, and interactions between these HA-liposome-coated surfaces were measured using a surface force balance (SFB) (*29*). The normal force profiles (Fig. 1A) show the onset of repulsion at ca. $D \approx 300$nm following addition of the liposomes, consistent with a monolayer of liposomes (diameter ca. 170 nm, fig. S1) attached to the PAH-HA coating on each surface.

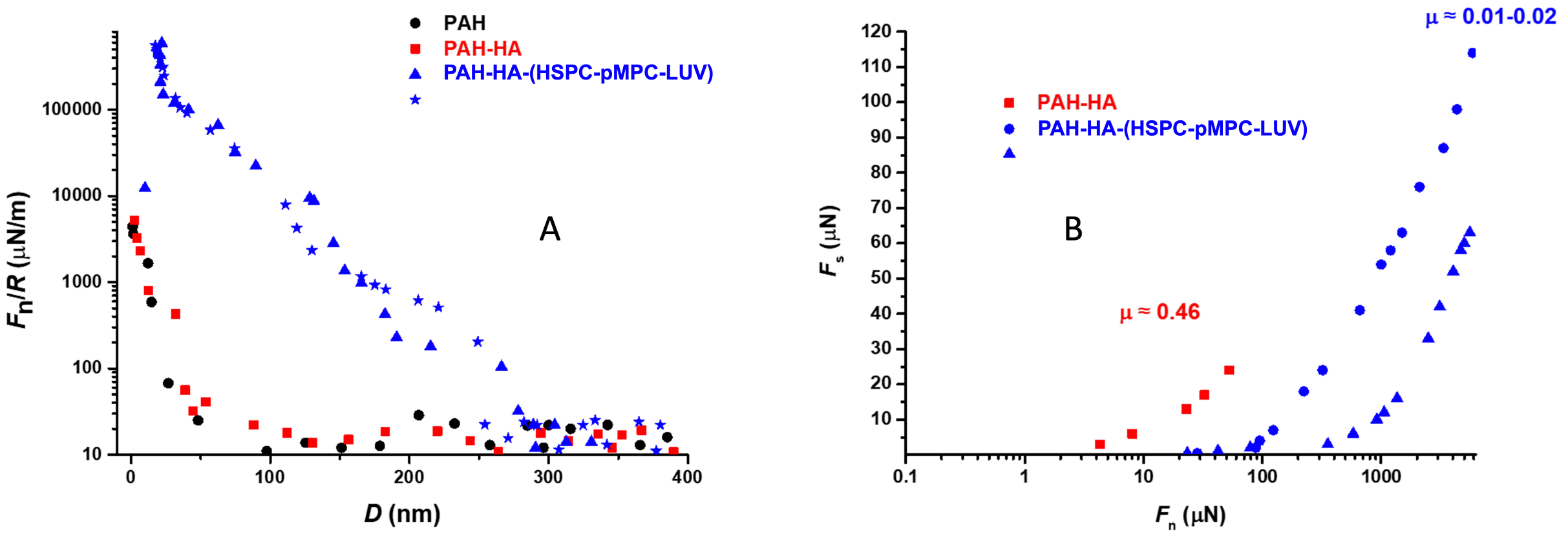

**Fig 1.** SFB measurements of pMPCylated-liposomes reveal their adsorption on cartilage-mimicking (HA-coated) surfaces across PBS ($NaNO_3$), and the associated strong friction reduction. (A) forces normalized by radius ($F_n/R$) as a function of surface separation ($D$) (B) friction forces ($F_s$) as a function of applied loads ($F_n$).

The shear-force vs. load profiles (Fig. 1B) reveal that friction is strongly reduced once liposomes are complexed with the surface-exposed HA, from μ ≈ 0.5 between PAH-HA coated mica (similar to the value μ ≈ 0.3 measured earlier between HA-coated mica surfaces across water(*32*)), to μ ≈ 0.01 – 0.02 once the lipids attach to the HA, up to contact pressures of ca. 4 MPa (comparable with physiological pressures in human joints(*1*)). This μ value, while higher than for pMPCylated HSPC-vesicles on bare mica(*36*) (possibly due to bridging effects(*37, 38*)), is similar to an earlier measurement of friction between mica-attached HA layers complexed with (non-functionalized) DPPC liposomes across salt solutions(*32*), for which μ ≈ 0.01. It is also comparable to values reported for friction of cartilage in vivo(*3*).

<u>Localization and retention of pMPCylated liposomes following injection into joints</u>

HSPC-pMPC-LUVs were fluorescently-labelled (DiR) and injected into the joints of naïve C57Bl/6 mice (fig. 2A) to examine their retention and localization (*29*). Fig. 2B shows the decay with time of the overall fluorescence intensity of the liposomes in the joint, revealing a retention half-life $t_{1/2} \approx 50$ hrs, somewhat shorter than the recently-reported retention time ($t_{1/2} \approx 85$ hrs) following IA injection of similarly-functionalized HSPC liposomes in a different murine model(*39*). Separately, to determine their localization on the cartilage and its time-variation, the liposomes were DiI-labelled and IA injected, following which the mice were sacrificed at different times, and the joints collected, cryo-embedded, sectioned and visualized using confocal microscopy (*29*) as shown in Fig. 3.

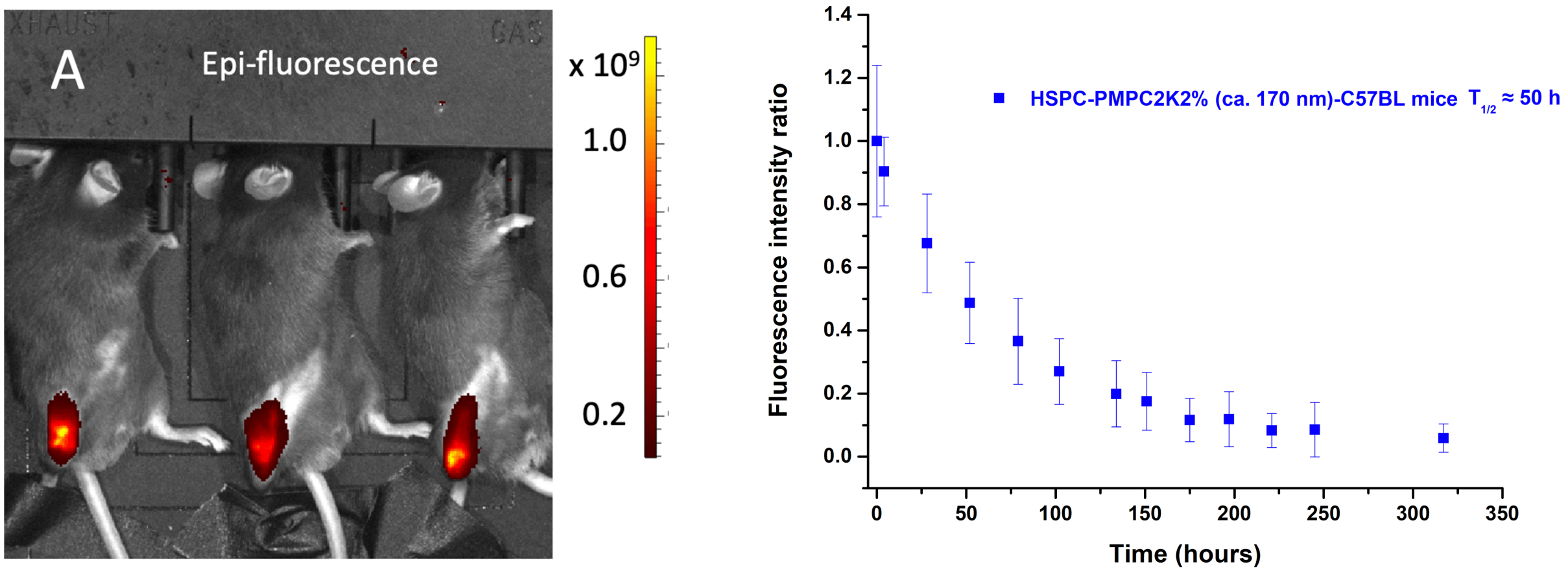

**Fig. 2.** Showing long retention time of pMPC-functionalized liposomes injected into naïve mouse joints. A) IVIS image of 3 repeats of fluorescence intensity for the liposomes injected into knee joints of C57BL/6 mice (at t = 153 hrs). Radiant efficiency in units of $(p/sec/cm^2/sr)/(mW/cm^2)$. B) Time decay of fluorescence intensity following IA injections, showing $t_{1/2}$ of ~50 hours. Mean diameter of the liposome vesicles (determined by DLS, fig. S1A) is ca. 170 nm.

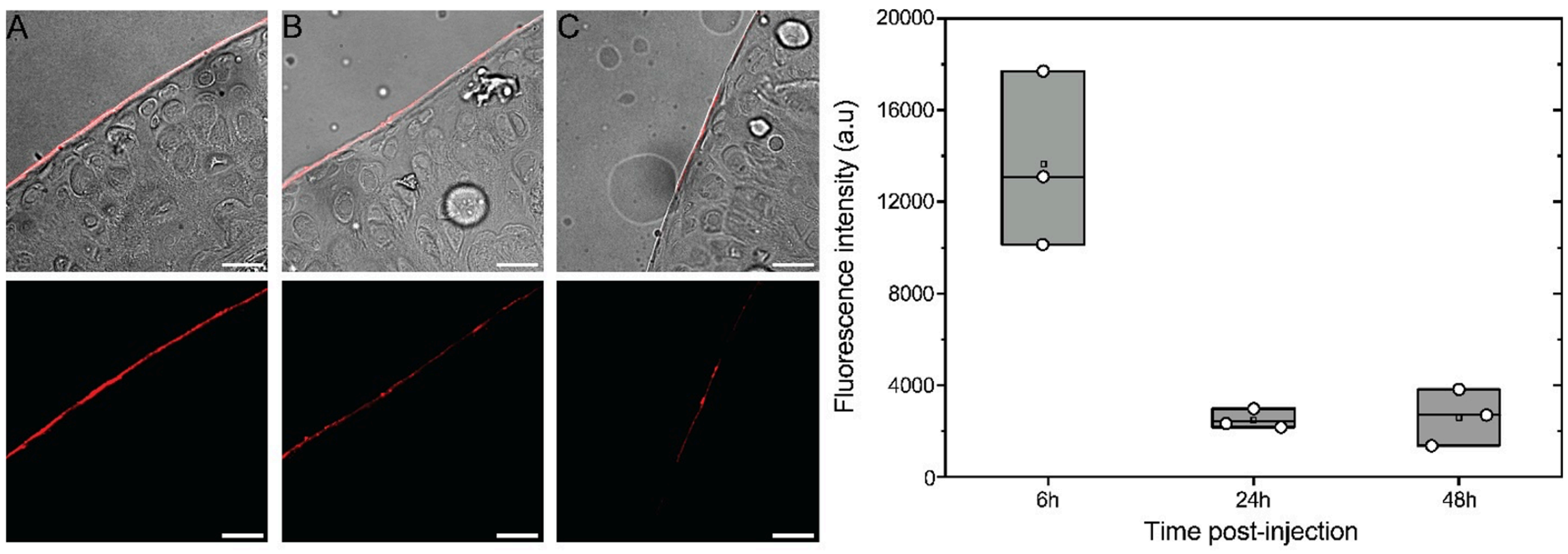

(i) (ii)

**Fig. 3.** pMPC liposomes injected into naïve mouse joints attach to the cartilage surface.
**i.** Representative confocal microscopy images of cross section of cryo-embedded mouse cartilage at (**A**) 6 h, (**B**) 24 h, (**C**) 48 h post knee injection with pMPC-liposomes incorporating DiI dye (0.5% mol/mol). Images are presented as fluorescent intensity of DiI signal (red) and overlay between liposomal fluorescence intensity and brightfield. Scale bar, 20 µm. **ii.** Retention profile of pMPC-liposomes stained with DiI dye (red). The data represents three independent biological repeats, with each data point an average of at least 10 spots on the cartilage surface for each biological sample. Boxes represent the 25-75 percentiles of the sample distribution (empty squares are mean of the data). Black horizontal lines represent the medians. All fluorescence microscopy data have been normalized to background by subtracting the values obtained for untreated cartilage.

Fig. 3 shows clearly that the HSPC-pMPC-LUVs attach to the articular surface and, while decreasing substantially by 24h, remain stable at this level on the cartilage at 48 hrs post-injection, indicating, in view of the fact that the cartilage surface is articulating under shear stress over that period, that they are robustly attached. Moreover, as seen in fig. 2, the liposomes are retained over many days at a lower level in the joint cavity (as also reported earlier in a different study(*39*)). It is of especial interest that both following 6 hrs, where it is clearly seen, but also following 24 and 48 hrs, the fluorescent signal is distributed on the cartilage surface either as a uniform continuous layer (6 hrs) or quasi-continuously (24, 48 hrs). In the latter case one would expect the lipids to spread and form a continuous layer on the surfaces when they slide past each other during articulation, as seen when discontinuous lipid micro-reservoirs at a hydrogel surface were slid past a countersurface(*40*). Since even a liposome monolayer(*41*) or a lipid bilayer(*40*) is sufficient to provide strong boundary lubrication, this implies that efficient lubrication should readily occur between the articulating cartilage surfaces at least up to 48 hrs post-injection.

Gene expression analysis following joint destabilization

Two days prior to joint destabilisation, 11-week male C57BL/6 mice animals received IA injections of either HSPC-pMPC-LUV dispersion or vehicle (PBS). Joints were surgically destabilised by destabilisation of the medial meniscus (DMM)(*42*), and animals were recovered from anaesthesia and allowed to mobilise normally (usually within 15 minutes). After 6h the animals were culled and whole joints extracted for RNA analysis of select mechano-sensitive genes (fig. S3A). From previous studies(*14*) these included genes considered to be shear responsive including *Mmp3, Il1b*, and non-shear responsive gene *Timp1.* We observed the increase of all three genes following DMM with PBS (fig. S3B). However, when considering the whole joint we did not see a difference in the expression level of genes between PBS and HSPC-pMPC-LUV treated groups (fig. S3B). As the whole joint contains multiple tissues, including cartilage, bone, meniscus, synovium as well as non-resident cells, e.g., inflammatory cells that infiltrate the joint in response to surgical injury, we decided to examine the articular cartilage separately. Accordingly, we repeated the previous experiment but collected only the cartilage tissue by microdissection for gene expression analysis (*29*) (Fig. 4A).

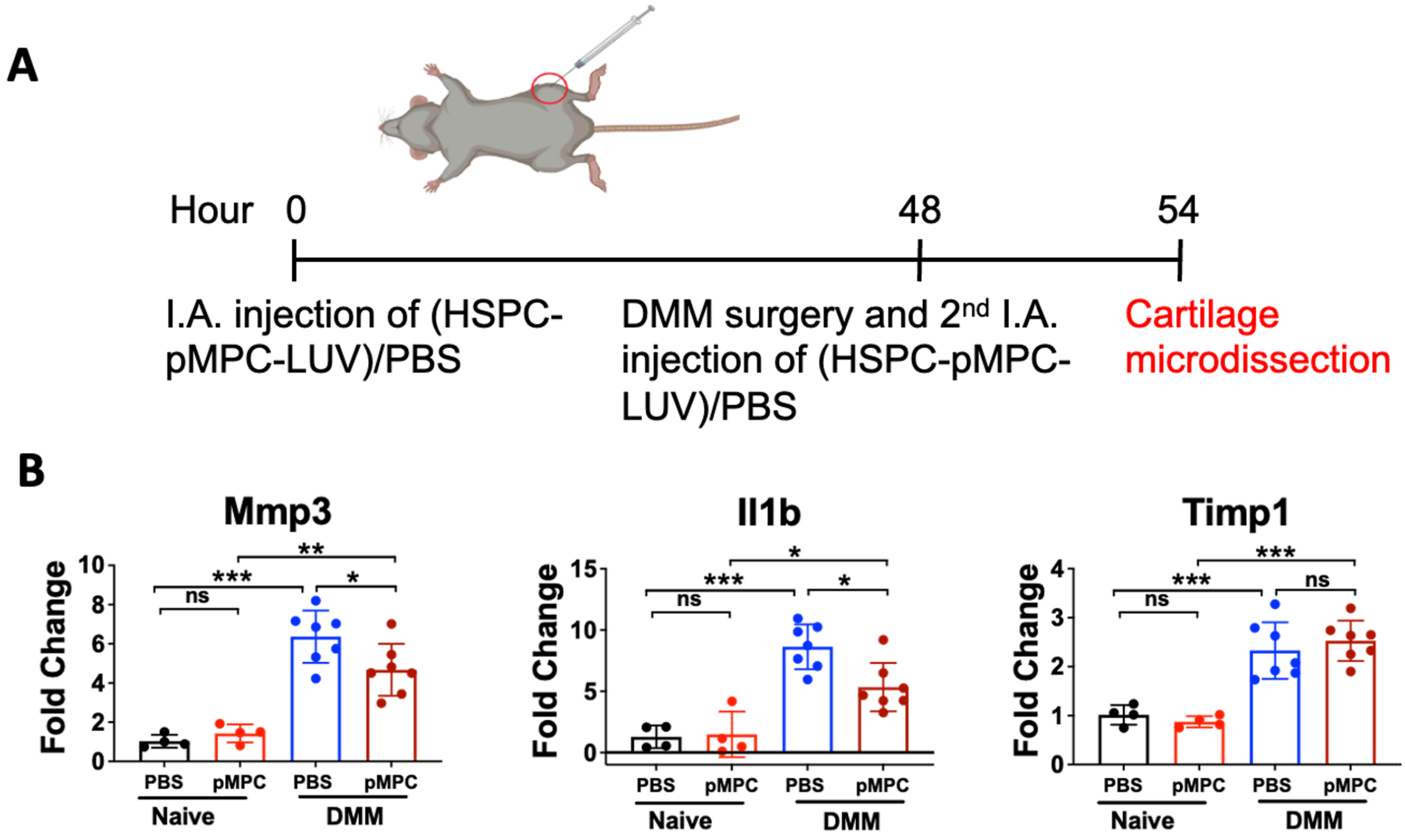

**Fig. 4.** Intra-articular injection of HSPC-pMPC-LUV lubricants suppresses shear-responsive genes in articular cartilage post DMM. **(**A) Schematic of the time course for the experiments. The right knee joints of male C57BL/6 mice were injected with 30 µl of 30 mM HSPC-pMPC-LUVs lubricant or PBS (vehicle) control. 48 hours later, at the time of DMM surgery, another IA injection of the lubricant or control was delivered. 6-hour post-surgery, cartilage was micro-dissected from knee joints and snap-frozen in RNAlater™. Cartilage from four joints was pooled together for RNA extraction for one data point. (B) Expression of shear-responsive genes *Mmp3* and *Il1b*, and non-shear responsive gene *Timp1* in the micro-dissected cartilage. Columns labelled 'pMPC' denote injection of the HSPC-pMPC-LUV lubricants. Error bars denote mean ± SEM. Results were expressed relative to *18s*, and p values were indicated for each gene using two-way ANOVA with Bonferroni's post hoc analysis. *p < 0.05, **p < 0.01, and ***p < 0.001.

DMM surgery upregulated the expression of *Mmp3* (fold-change 6.36 ±2.24)*, Il1b* (8.63 ± 3.33) and *Timp1* (2.33±0.52) in the articular cartilage compared with naïve animals (Fig. 4B), but in the presence of the lubricant the increase in shear responsive genes *Mmp3* and *Il1b* was suppressed (down to *Mmp3*, 4.67±1.65, and *Il1b* 5.34 ± 2.26), while the expression level of the non-shear responsive gene, *Timp1,* was not affected (2.53 ± 0.43) (Fig. 4B). In a parallel study (*29*) the surgically destabilised joint was immobilised immediately after surgery with a thermoplastic splint which allowed weight bearing (joint compression) but prevented knee flexion and hence cartilage surface shear stress (fig S4A). After 6h, mice were culled and the joints were collected for qRT-PCR as before, showing that, similar to the effect of lubricant addition, such immobilisation suppressed shear associated genes *Mmp3* and *Il1b*, but had no effect on compression activated gene, *Timp1* (fig S4B). This result corroborates our finding that intra-articular injection of HSPC-pMPC-LUVs was able to change the biological consequences of joint destabilisation by reducing shear stress on cartilage.

Direct visualization (RNAscope®) of gene expression in cartilage-embedded chondrocytes

RNAscope® (ACD Bio) is a powerful tool (*29*) for sensitive detection of mRNA transcripts at high resolution that allows one to visualize the expression of specific genes *in situ*. We repeated the experiment above, save that sham-operated mice (in which the joint was opened but not destabilised) were compared with DMM-operated animals. Each group either had the HSPC-pMPC-LUV or PBS injections according to Fig. 5A. To overcome the challenges of performing RNAscope on whole, non-decalcified murine joint sections, we adapted a modified protocol (Fig. 5A)(*43*). After applying an increased sucrose gradient on the joints, the tissue was sectioned coronally using a cryostat. Adjacent sections from each joint were processed for safranin-O staining and *in situ* hybridization (using the RNAscope Multiplex Fluorescent v2 system). Safranin-O-stained sections (*29*) were necessary to confirm the integrity of the section after processing and were used to delineate the boundary between cartilage and subchondral bone (Fig.5B). RNAscope images were acquired from the medial tibial plateau, the area which degrades first after joint destabilisation and where shear stresses are predicted to be greatest(*44*). Images are shown for *Mmp3* and *Timp1* (Fig. 5C) with quantification by counting the percentage of cells showing a positive signal (Fig. 5D). The number of positive cells for *Mmp3* in the DMM-PBS samples (25.9 ± 5.2%) increased more than two-fold compared with the sham-PBS samples (10.3 ± 3.8%). Injection of HSPC-pMPC-LUVs suppressed the percentage of *Mmp3* positive cells back to the same level as the sham-operated samples (10.6 ± 4.8%) (Fig. 5D). In contrast, there was no difference in the percentage of the positive cells for the expression of *Timp1* in all four groups, indicating that HSPC-pMPC-LUVs only inhibited the injury-induced shear-responsive genes (Fig. 5D). Importantly, these responses were observed mostly in the superficial chondrocytes, 0-20µm from the articular surface (2-3 cell layers deep) (Fig. 5E); this is expected because the shear stress on the chondrocytes due to friction at the cartilage surface decays with tissue depth, as later shown.

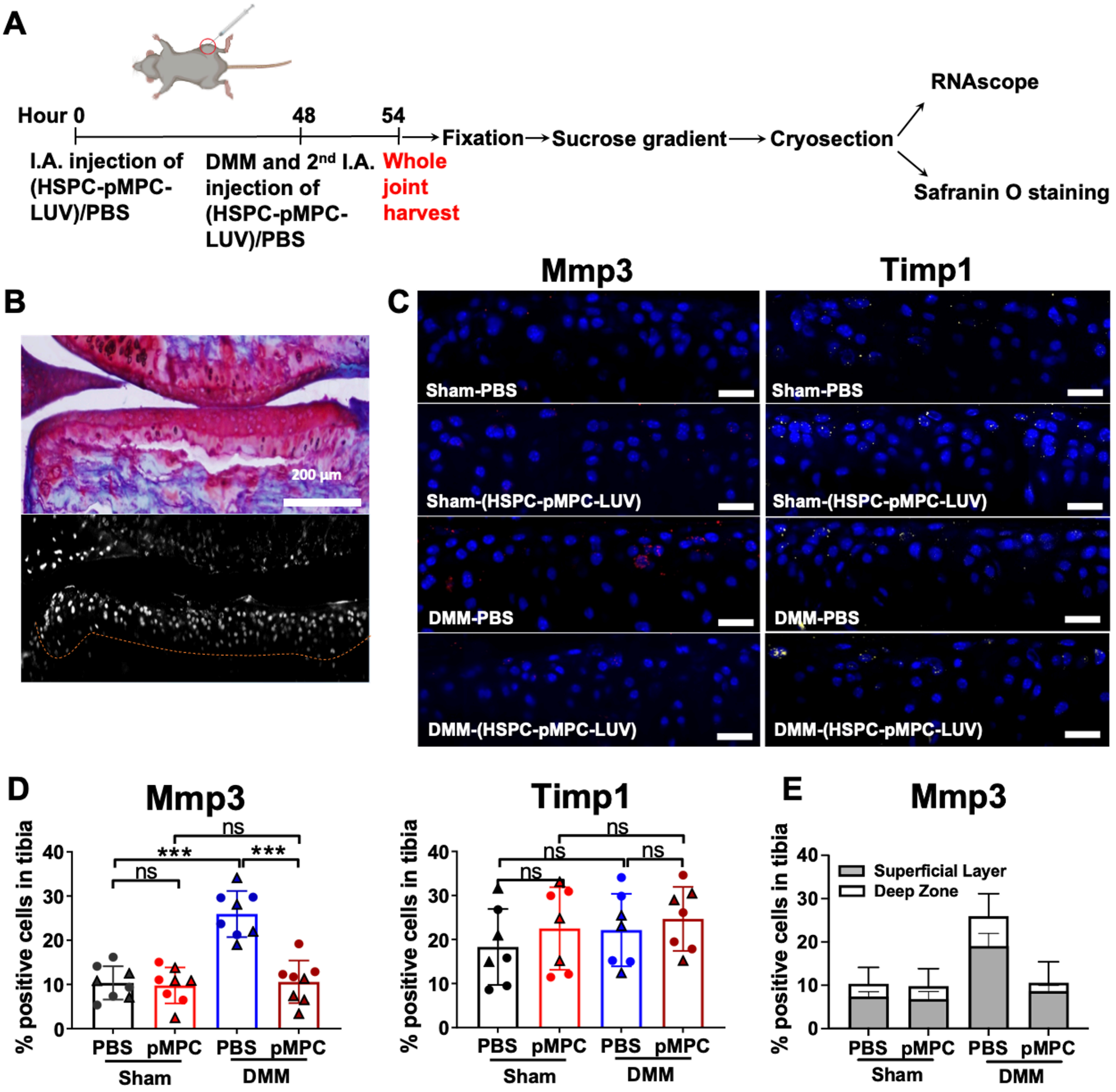

**Fig. 5.** Visualizing the expression of mechano-responsive genes in articular cartilage by RNAscope®. (A) Schematic showing the study protocol of the RNAscope and safranin O staining. Mouse whole knee joints were harvested after two doses of intra-articular injection and DMM surgery as before (Fig. 4A) and fixed in 4% ice-cold paraformaldehyde (in PBS) for 24 hours. Joints were subjected to an increasing sucrose gradient and sectioned by cryostat (8 μm). Adjacent sections were used for RNAscope and safranin O staining. (B) Representative image of safranin O staining (top) and dapi signals (bottom) of adjacent sections from the same knee joint. (C) mRNA transcripts of *Mmp3* (red dots) and *Timp1* (yellow dots) in cartilage from the medial tibia plateau in Sham-PBS, Sham-(HSPC-pMPC-LUV), DMM-PBS and DMM-(HSPC-pMPC-LUV) treated mice. Scale bar, 20 μm. (D) Quantification of the *Mmp3*- and *Timp1*-positive cells in tibia cartilage from (C). (E) Distribution of the *Mmp3*-positive cells within the superficial (top two layers of chondrocytes) and deep zones of the tibia cartilage from (C). Columns labelled 'pMPC' denote injection of the HSPC-pMPC-LUV lubricants. p values are indicated for each gene using two-way ANOVA with Bonferroni's post hoc analysis. $^{**}p < 0.01$, and $^{***}p < 0.001$. n= 8, solid circles and empty triangle represent two batches of experiments.

## Discussion

The main new finding of this study is that injecting poly(phosphocholinated) liposomic lubricants (HSPC-pMPC-LUVs) into joints of mice that have been surgically destabilized (DMM), suppresses upregulation of shear-stress-sensitive mechano-inflammatory genes including inflammatory cytokines and matrix-degrading enzymes. At the same time the regulation of a mechanosensitive gene that is non-shear-responsive was unaffected by the injection of the lubricant. These results were corroborated by immobilising the mouse joint using a splint, which reduces joint shear stress by preventing its articulation, but allows compressive load as the mice still bear weight through this limb. Our findings moreover accord with previous work where sciatic neurectomy, which likewise suppresses joint articulation (due to paralysis of the knee flexors), also prevented inflammatory gene regulation whilst preserving compression associated genes. The pMPCylated liposomes used as lubricants, which are known to act as excellent boundary lubricants in model studies(*36*), are shown to massively reduce friction also when attached (physically rather than chemically, likely through an entropy-based counterion release mechanism(*45*)) to an HA-coated substrate mimicking HA-rich articular cartilage surfaces(*1, 34, 35*). These liposomic lubricants are also shown to have long in vivo retention time in mouse joints in our study (Fig. 2), in line with earlier work(*39*). Importantly, they persist as either continuous or quasi-continuous adsorbed layers on the cartilage surface in joints in vivo (Fig. 3), over at least the duration of our study (ca. 2 days).

Together these results point clearly to the following scenario: the IA-injected liposomes form lubricating layers on the articular cartilage in the DMM-operated mice. These result in a reduced friction coefficient $\mu$ as the joint articulates, where the surface shear stress $\sigma_s$ is given by $\sigma_s(z)_{z=0} = \mu P$ at the articular cartilage surface (where P is the contact pressure between the cartilage surfaces and z the cartilage depth, where z = 0 denotes the cartilage surface). The effect is to reduce the shear stress throughout the cartilage, and suppress the regulation of shear-stress-sensitive genes, including *Mmp3* and *Il1b,* relative to the unlubricated DMM-operated joint (Figs. 4 and 5).

This scenario is further supported when one considers the shear-stress $\sigma_{s,chondrocyte}(z)$ experienced by the chondrocytes at different depths z within the cartilage. Even though it is expected, on general grounds of stress balance, that the shear stress within the cartilage layer itself is essentially independent of the depth z and is given throughout by its value at the cartilage surface (eq 1), this, counterintuitively, is not the case for the shear stress on the chondrocytes. This is because, as is well established, the modulus of articular cartilage $K_{cartilage}$ is lower near its surface (superficial zone) than it is deeper down(*46, 47*), while at the same time the typical rigidity modulus of cells $K_{cell}$ – including chondrocytes – is far lower than $K_{cartilage}$. Given this it is readily shown (*29*) that $\sigma_{s,chondrocyte}(z)$ is given by

$$\sigma_{s,chondrocyte}(z) = \mu P[K_{cell}/K_{cartilage}(z)] \quad (1)$$

Since the rigidity modulus $K_{cartilage}(z)$ increases with depth(*47*), eq 2 predicts that the shear stress on the chondrocytes *decreases* with depth. We would therefore expect to see more shear-stress-sensitive genes expressed near the cartilage surface (in the superficial layer) than deeper down. That is exactly what the RNAscope results show, Fig. 5E, where most of the *Mmp3*-positive cells are indeed within the superficial zone. This further supports the concept that it is the lubrication by the HSPC-pMPC-LUVs that is directly responsible for the observed suppression of shear-responsive genes.

It is worth noting that RNAscope signals were generally low in our samples compared with published studies using other tissues(*43*). This perhaps reflects the chondrocytes' relatively 'quiescent' phenotype or added complexities due to cartilage's highly charged, matrix-rich tissue. This may have contributed to the inability to detect robust regulation of certain genes such as the lack of regulation of *Timp1* on RNAscope after joint destabilisation, even though this was readily observed by RT-PCR when cartilage was microdissected (fig. 4B). Few of the RNAscope probes have been validated in cartilage previously and additional refinements are likely to be needed in future studies.

In sum, we demonstrate for the first time the relation between cartilage lubrication in vivo and the regulation of shear-stress-sensitive genes in chondrocytes. This is achieved through IA administration of pMPCylated HSPC liposomes which coat the cartilage surface and act, purely though hydration lubrication at the cartilage surface, with no chemical interactions involved, to reduce the friction and thus the shear stress transmitted to the embedded chondrocytes. In view of the fact that shear-stress-sensitive pathways are linked to OA progression(*12, 13*), our results suggest novel treatment modalities, based on IA administration of exogenous liposome-based lubricating agents, for alleviating OA or slowing its progression. Future studies will explore the longer term effects of lubricant dosing in vivo on development of experimental OA.

**Acknowledgements**

We thank Sam Safran for useful comments on the ms. We thank the European Research Council (Advanced Grant CartiLube 743016), the McCutchen Foundation, the Israel Ministry of Science and Technology (Grant 3-15716), and the Israel Science Foundation (Grant 1229/20) for financial support. This work was made possible partly through the historic generosity of the Perlman family.

**Competing interests**

The Weizmann Institute has a patent on polyphosphocholinated–lipid conjugates and liposomes stabilized by such conjugates (US10730976B2), on which WL, RG and JK are co-inventors. RG is a founder of Liposphere, founded 2019; her involvement in this project was prior to this time.

**Supplementary Materials**

Materials and Methods

Figs. S1 to S4

References (*48 – 49*)

## Materials and Methods

Animals

Ten-week-old male C57BL/6 mice were purchased either from Jackson Laboratory (retention and cartilage adsorption studies at the Weizmann Institute) or from Charles River (for all other experiments at the Kennedy Institute of Rheumatology (KIR) in Oxford). Mice at the Weizmann Institute were housed in groups of 1–5 mice per micro-isolator cage in a room with a 12 h light/dark schedule with free access to food and water. Mice at the KIR were kept in an approved animal-care facility and maintained under 12-hour light/12-hour dark conditions at an ambient temperature of 21°C; housed 4 per cage in standard, individually ventilated cages and were fed with certified mouse diet (RM3; Special Dietary Systems) and water ad libitum. All animal experiments were performed according to the guidelines established by the Weizmann Institute of Science, Rehovot, Israel, and approved by the Institutional Animal Care and Use Committee (IACUC), or following ethical and statutory approval in accordance with local policy at the University of Oxford.

Materials

HSPC (Hydrogenated soybean phosphatidylcholine, Mw 786.11) was purchased from Lipoid GmbH (Germany). DiI Stain (1,10-dioctadecyl-3,3,30,30-etramethylindocarbocyanine perchlorate C59H97ClN2O4, Mw 933.88) and DiR ((1,1'-Dioctadecyl-3,3,3',3'-Tetramethylindotricarbocyanine Iodide, C63H101IN2, Mw 1013.40) and Pierce™ Rapid Gel Clot Endotoxin Assay Kit were provided by ThermoFisher Scientific (Waltham, MA, USA), Cell Proliferation Kit (XTT based) was purchased from Biological Industries (Israel Beit Haemek LTD, Israel). Holey carbon grid (C flat 3/2 200 mesh) was purchased from Electron Microscopy Science (Hatfield, PA, USA). DSPE–pMPC2k (molecular weight of pMPC is ca. 2 kDa) was prepared by two steps according to the reported procedure(*36*) with minor adjustment, for the subsequent preparation of the HSPC-pMPC liposomes. Poly(allylamine hydrochloride) (PAH, Mw 50000) was purchased from Sigma-Aldrich (Israel). Hyaluronic acid (1.5 M) was purchased from CreativePEG (North Carolina, USA). For all IA administration in the in vivo experiments, 33 mM of the HSPC-pMPC liposomes were used.

Preparation of large unilamellar vesicles (LUVs)

HSPC (98% mol/mol) / pMPC (2 kDa, 2% mol/mol) / 0.8% DiI (mol/mol) [for cartilage adsorption studies] or DiR (0.1% mol/mol) [for cartilage retention studies] were dissolved in chloroform/methanol mixture (2:1) and organic solvent was evaporated using first nitrogen stream, followed by 8 h of vacuum pumping. The lipid film was then hydrated with PBS solution (osmol = 315 mOsmo/kg) at 70˚C to reach the desired concentration and solution was gently vortexed. The resulting MLV suspensions were sonicated for 15 min at 70˚C to disperse larger aggregates. The vesicles were subsequently downsized by extrusion (Lipex, Northren Lipids Inc.) through 400 nm and 200 nm polycarbonate membrane. The extrusion was performed 11 times through each membrane at 65˚C. To remove small size liposomes and to uniform the sample, liposomes were centrifuged at 15000 rpm for 90 min. Supernatant was discarded and pallets were resuspended in appropriate volume of PBS. Liposomal samples were tested for endotoxin content using the Pierce LAL chromogenic LPS test (Thermo Fisher Scientific, Paisley, UK) following the manufacturer's instructions.

<u>Evaluation of pMPC-liposomes cytotoxicity</u>

Liposomes cytotoxicity was determined by the production of the yellow formazan product upon cleavage of XTT by mitochondrial dehydrogenases in viable VERO cells (kidney epithetical cell line derived from an African Green Monkey). The cells were seeded onto 96-well plates (~ $4\times10^4$ cells/well) in RPMI media (+ 10% FBS + 1% pen/strep). When a confluent state was reached (usually after 24 h), 50 µL of pMPC liposomes (at concentrations: 1.0 mM, 0.5 mM, 0.25 mM and 0.125 mM) in PBS solutions were then added. After 24 h, the cells were incubated with 50 µL of XTT solution composed for 3h. Absorbance values were later measured with a multiwell-plate reader (Cary 100 Bio, Varian Inc, USA) at a wavelength of 450 nm. Background absorbance was measured at 620 nm and subtracted from the 450 nm measurement. The experiments were repeated at least three times for each liposomes concentration. RPMI culture medium was used as a positive control.

The potential toxic effect of the pMPC liposomes tested was expressed as a viability percentage calculated using the following formula:

$$\%Viability=100-[(ODtest/ODc)\times100]$$

Where ODtest was the optical density of those wells treated with the liposome solutions, and ODc was the optical density of those wells treated with supplement-free RPMI media.

Dynamic light scattering measurements

The size and zeta-potential of the lipid nanoparticles were measured with a ZetaSizer Nano ZS (Malvern Instruments, UK) at 25˚C. Triplicate measurements with a minimum of 10 runs were performed for each sample.

Cryo-TEM measurements

A Vitrobot Mark IV plunger system was used to prepare the samples for cryo-TEM. Humidity was kept close to 80% for all experiments and the temperature was set at 24 ˚C. 3.5 µL of the sample were placed onto a holey carbon grid (C flat 3/2 200 mesh) which was rendered hydrophilic via glow discharge (Evactron Plasma Cleaning, Evactron, USA). Excess sample was removed by blotting with filter paper and the sample grid was vitrified by rapid plunging into liquid ethane (-180 ˚C). The grids were kept in liquid nitrogen before being transferred into a Gatan 626 cryo-holder. Cryo-TEM imaging was performed on a FEI Tecnai T12 TEM (120 kV) with a with a TVIPS F244HD CCD digital camera.

Cartilage adsorption studies

Liposomes suspension (10 µL, 15 mM) was administered intraarticularly (IA) into right knee using BD Micro Fine (30G) syringes under Isoflurane anesthesia. Buprenorphine (10 µL, 0.3mg/mL) was intraperitoneal (IP) injected using BD Micro Fine (30G) syringes to relieve pain caused by IA injection. At 6 h, 24 h and 48 h post injections mice were euthanized in their home cages by exposure to carbon dioxide, and joints were collected.

Cryo- embedding, sectioning of samples and visualization

Samples were fixed in 4% PFA for 48 hours following decalcification for another 48 hours in Calci-Clear rapid decalcifying agent (National diagnostics, USA). Samples were then incubated in 30% sucrose solution and embedded in cryo-molds using OCT compound (Tissue-Tek, USA). Seven microns thick coronal section through the middle of the knee joint were prepared using the Leica CM1950 cryotome (Leica Biosystems Newcastle Ltd, UK). Slides were air-dried for 30 minutes following mounting with DAPI-containing mounting media (Fluoromount, BSB-0163 by BioSB USA). Confocal pictures were acquired using a confocal laser scanning fluorescence microscope LSM700 (Zeiss) (Olympus Life Science, Olympus Co., Japan). All images were acquired using a 60X, oil immersion objective, and individual field of view were subsequently stitched together to form the full section. Images were recorded in brightfield mode and in confocal mode using a 561 excitation laser channel. Picture analysis was performed using ImageJ software v1.52i (NIH, USA). For comparative analysis, all parameters during image acquisition were kept constant throughout each experiment.

Intra-articular retention studies

Mice received IA injections of the pMPC liposomes (10 µL, 15 mM) to the right joint space using a BD Micro Fine (30G) syringe under isoflurane anesthesia. The hair around the hind limb surgical sites was removed. Mice were scanned in an IVIS imaging system (PerkinElmer Inc., USA) at different time points. The excitation and emission detectors were set at 760 nm and 780 nm, respectively.

Surface force balance (SFB) measurements

The SFB measures directly the normal and shear forces between atomically-smooth mica surfaces (either bare or coated with other species) and is described extensively in previous work, e.g. refs.(*25, 26, 36*). The mica surfaces were incubated in PAH solution (0.1 mg/mL) in 0.15 M PBS (0.14 M $NaNO_3$) for 5 mins, and then rinsed with 0.15 M PBS (0.14 M $NaNO_3$). Hyaluronic acid was dissolved in 0.15 M PBS (0.14 M $NaNO_3$) at 0.10 mg/mL, the solutions were stirred overnight before use. Then, the mica-PAH surface is dipped in a 0.10 mg/mL HA solution. After overnight incubation, the excess HA molecules on the surfaces were rinsed in a beaker containing 200 mL of 0.15 M PBS (0.14 M NaNO3). The mica-PAH-HA surfaces were immersed overnight in the HSPC/pMPC2k liposome dispersions (0.3 mM) added to the SFB bath, and the subsequent SFB experiments were carried out in the liposome dispersions. The mean contact pressure P between the compressed mica surfaces of unperturbed radius of curvature R under a load $F_n$ may be estimated as $P = F_n/(\pi a^2)$ from the Hertzian expression for the contact radius $a \approx (F_nR/K)^{1/3}$, where K is the effective modulus of the mica/glue layer on the crossed cylindrical glass mounts in the SFB. For $R \approx 0.01$ m and $K \approx 5.10^9$ N/m$^2$, (refs. 1 - 3) this yields a pressure of $P \approx 4$ MPa at the highest values of $F_n$ (fig. 1).

Surgical procedures.

Male mice (ages 11-12 weeks) undergoing surgery were anesthetized by inhalation of isoflurane (3% induction and 1.5-2% maintenance) in 1.5-2 L/min oxygen. All mice received a subcutaneous injection of buprenorphine (Vetergesic; Alstoe Animal Health) after surgery. The mice were fully mobile within 4-5 min after the withdrawal of isoflurane. Induction of OA by DMM was performed as previously described.(*12*) Mice used as sham controls were anesthetized as for DMM. The right knee was opened using the same medial parapatellar approach to identify the meniscus without releasing the meniscotibial ligament. The first dose of HSPC-pMPC SUVs (33 mM, 30 µL) or PBS was injected into the right knee synovial cavity two-days before surgery (DMM, Sham-operated or Naïve mice) under anaesthetic. The second dose of HSPC-pMPC SUVs or PBS was injected in the joint right after the surgery (DMM or Sham-operated mice) or 6 hours before tissue harvesting (naïve mice).

Mouse joint splints:

Computer aided design was used to create a splint design based on CT scans of a mouse hind limb. The splints were 3D printed in polylactic acid. Following surgery, the splint was slipped over the hind limb to immobilise the knee. The ankle and foot were free to move. Once in position silicone was injected via a side portal in the splint. The liquid silicone filled any gaps between the splint and the limb. A window in the splint over the anterior knee provided access to the wound, which was left uncovered by the silicone. The splint was thin (< 0.75 mm) and light (< 0.8g). In pilot studies, up to seven days, the splints were well-tolerated, activity levels maintained and did not cause skin abrasion.

RNA extraction and real-time quantitative reverse transcription-polymerase chain reaction (qPCR).

6-hour post-surgery, mice were culled by $CO_2$. Whole joints were collected and snap-frozen in liquid nitrogen, and then stored at -80 $^0$C. RNA was extraction from the joint using the RNeasy Mini Kit (Qiagen) according to the manufacturer's instructions. Articular cartilage was collected by micro-dissection and kept in RNA later (Invitrogen™, AM7020). Cartilage from four mice was pooled together for RNA isolation to get one data point. Complementary DNA (cDNA) was generated from RNA using a High Capacity cDNA kit (Applied Biosystems) following the manufacturer's instructions. qPCR was run on custom-designed TLDA microfluidic cards ordered from Applied Biosystems. All thermocycling was performed on the ViiA™ 7 system (Applied Biosystems). 50 µl of TaqMan Universal PCR Master Mix (Applied Biosystems) were mixed with 50 µl cDNA template in nuclease-free water and added to each TLDA loading port. The TLDA card was then centrifuged and sealed. Ct values were obtained after manually choosing the Ct threshold. Expression of the respective genes was normalized to that of *18s* as an internal control, using the $2^{-\Delta\Delta Ct}$ method.

In Situ Hybridization and Safo-O staining.

Freshly collected whole joints were fixed in 4% (in PBS) freshly prepared ice-cold paraformaldehyde (PFA) for 24 h at 4°C. After washed in ice-cold PBS three times, the joints were applied with increasing sucrose gradient at 4°C: 10%, 20% and 30% sucrose in PBS for 24 hours each gradient. Then joints were quickly frozen in Super Cryo Embedding Medium (Section lab, Japan) over dry ice and stored at −80 C° until sectioning on a cryostat. 8 µm thickness of coronal section through the middle of the knee joint were collected with Kawamoto's tape (Cryofilm type 3C(16UF), 2 or 2.5 cm). Kawamoto's tape with a section was then fixed on polysine slide (Fisher brand Superfrost Plus, Fisher Scientific) using Tough-Spots tap (Sigma-aldrich). The slides were air-dried inside the cryotome at -20°C and then kept at −80°C to preserve high RNA integrity.

Prior to hybridization, sections were rinsed with PBS and baked at 60 °C for 30 mins and then dehydrated in increasing concentrations of ethanol (50%, 70%, 95%, 100%, 100%). Sections were incubated with probes for 2 hours (*Mmp3*, ACD Cat #. 480961, *Timp1,* ACD Cat # 316841). RNAscope Multiplex Fluorescent Detection kit v2 (ACD, Cat #: 323100) were used to amplify the probe according to the manufacturer's instructions. Opal™ dye (620 Reagent Pack, Akoya Biosciences, Cat # FP1495001KT, 1/1000 dilution) was used for the fluorescence signals as the manufacturer recommended. Slides were coverslipped using ProLong® Gold Antifade Reagent with DAPI (Cell Signaling) and sealed with nail polish.

Adjacent sections were stained with haematoxylin, fast green and safranin O according to a modified protocol to help distinguish the chondrocytes from subchondral bone(*48*).

Imaging

RNAscope signals were acquired at single-cell resolution (frame size: 4096 × 4096 pixels; pixel size: 0.69 µm) on a Zeiss Laser Scanning Microscope (LSM) 880 equipped with seven laser lines (405, 453, 488, 514, 561, 594 and 633 nm). Three tiles covered the whole medial tibia cartilage surface were taken for each joint with a magnification of 40 ×. Safo-O staining images were taken on an Olympus BX51 Ostometric fluorescence microscope.

Transcript quantification in Chondrocytes

Imaris Cell Imaging Software (version 9.0) were used to quantify the transcripts per chondrocytes with supervision by two independent blinded scorers. Quantification of the transcripts was performed using segmentation analysis with a minimum threshold diameter of 0.4μm and 4.0μm for RNA signal and DAPI respectively. Segmentation of the articular cartilage and superficial cartilage were manually drawn with reference to consecutive sections stained using Safranin O. Colocalization analysis was performed for RNA positive cells within the segmented regions of interest.

Shear stress on cartilage-embedded chondrocytes

While we expect, on general grounds of stress balance, that the shear stress within the cartilage layer itself is essentially independent of its depth z and is given throughout by its value

$$\sigma_s(z) = \sigma_s(z)_{z=0} = \mu P \quad (1)$$

where P is the contact pressure at the cartilage surface (z = 0), this is not the case for the shear *strain*. The shear strain $\varepsilon(z)$ at any depth z within the cartilage is given by the usual relation $\varepsilon(z) = [\sigma_s(z)/K_{cartilage}(z)]$ where $K_{cartilage}(z)$ is the local rigidity- or shear-modulus of the cartilage at that depth. $\varepsilon(z)$ varies with depth because, as is well established, the modulus $K_{cartilage}$ is lower near its surface (superficial zone) than it is deeper down(*46, 47*). Since the typical rigidity modulus of cells $K_{cell}$ – including chondrocytes – is far lower than that of articular cartilage (for which $K_{cartilage} = O(10^5 – 10^6$ Pa)(*46, 47*), with values $K_{cell} = O(100$ Pa)(*49*)), and since the volume fraction of chondrocytes in the cartilage is low (1 – 3%), the strain deformation $\varepsilon_{cell}$ of the chondrocytes will be determined by the local strain $\varepsilon(z)$ in the cartilage, i.e.

$$\varepsilon_{cell} = \varepsilon(z) = [\sigma_s(z)_{z=0}/K_{cartilage}(z)] \quad (2)$$

The corresponding shear stress $\sigma_{cell}(z)$ experienced by the chondrocytes is therefore, from eqs (1) and (2), given by

$$\sigma_{cell}(z) = (K_{cell} \text{ x } \varepsilon_{cell}(z)) = [K_{cell} \text{ x } \sigma_s(z)_{z=0}]/K_{cart}(z) = \mu P[K_{cell}/K_{cart}(z)] \quad (3)$$

which is eq 1 in the main text.

## Supplementary Figures and description

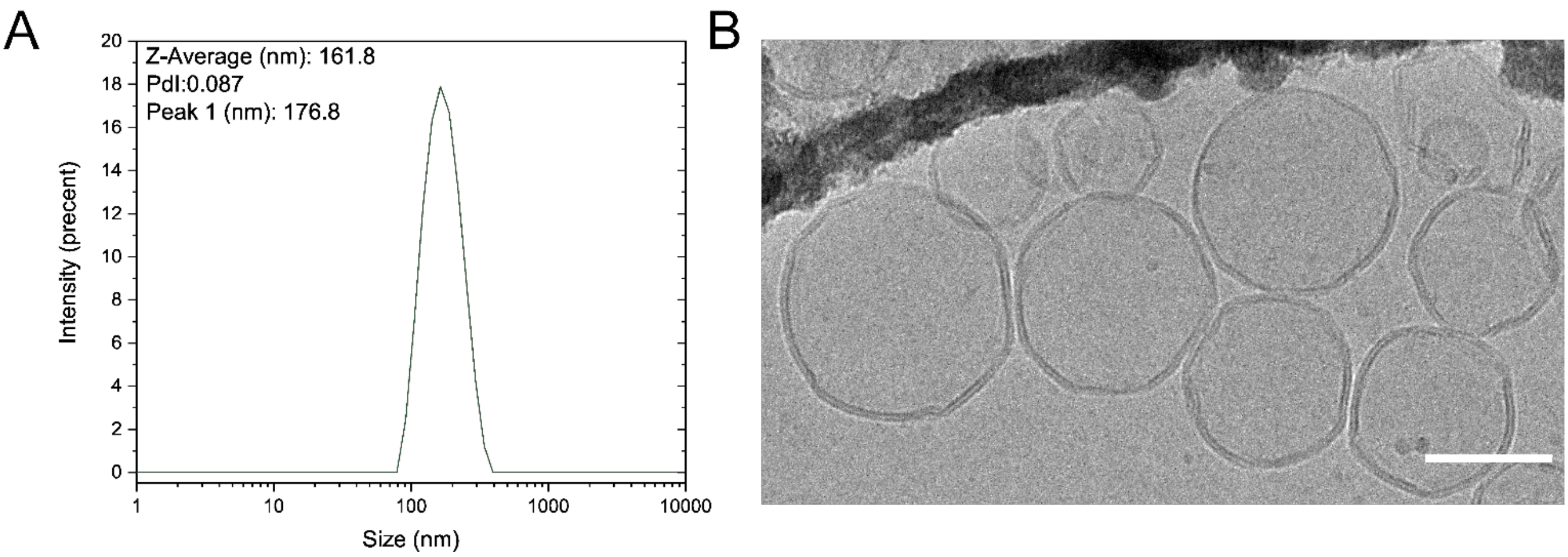

**Fig. S1.** Characterization of pMPCylated-liposomes prepared in PBS by extrusion through 200 nm **(A)** Size distribution and **(B)** Representative cryogenic electron micrograph of pMPC-functionalized liposomes used in the study. Scale bar, 100 nm

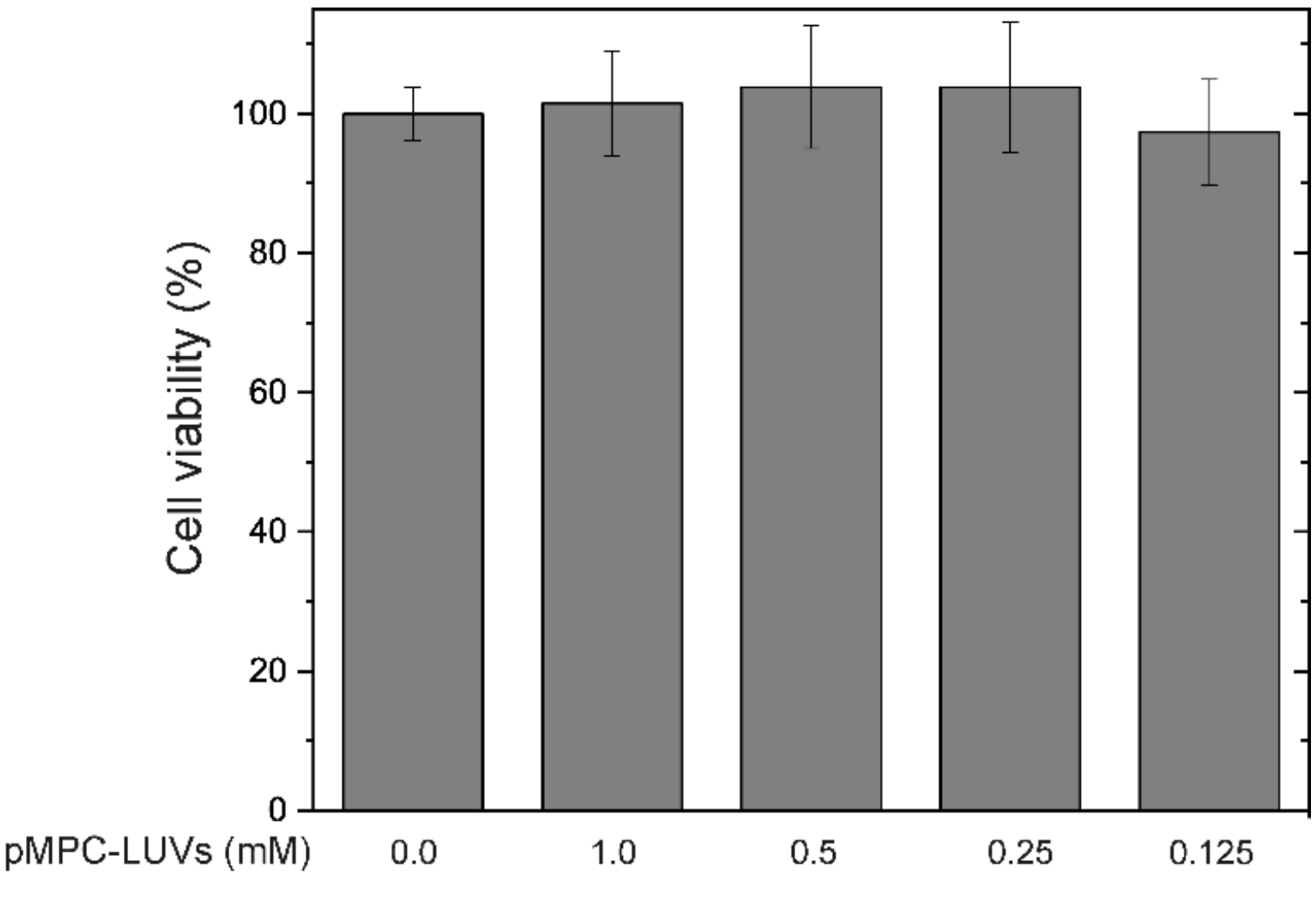

**Fig. S2.** HSPC-pMPC-LUVs are non-toxic to VERO cells. Cell viability of VERO cells upon 24 h-incubation with pMPC-functionalized liposomes at various concentrations. The data represents the averages and standard deviations from three independent biological repeats.

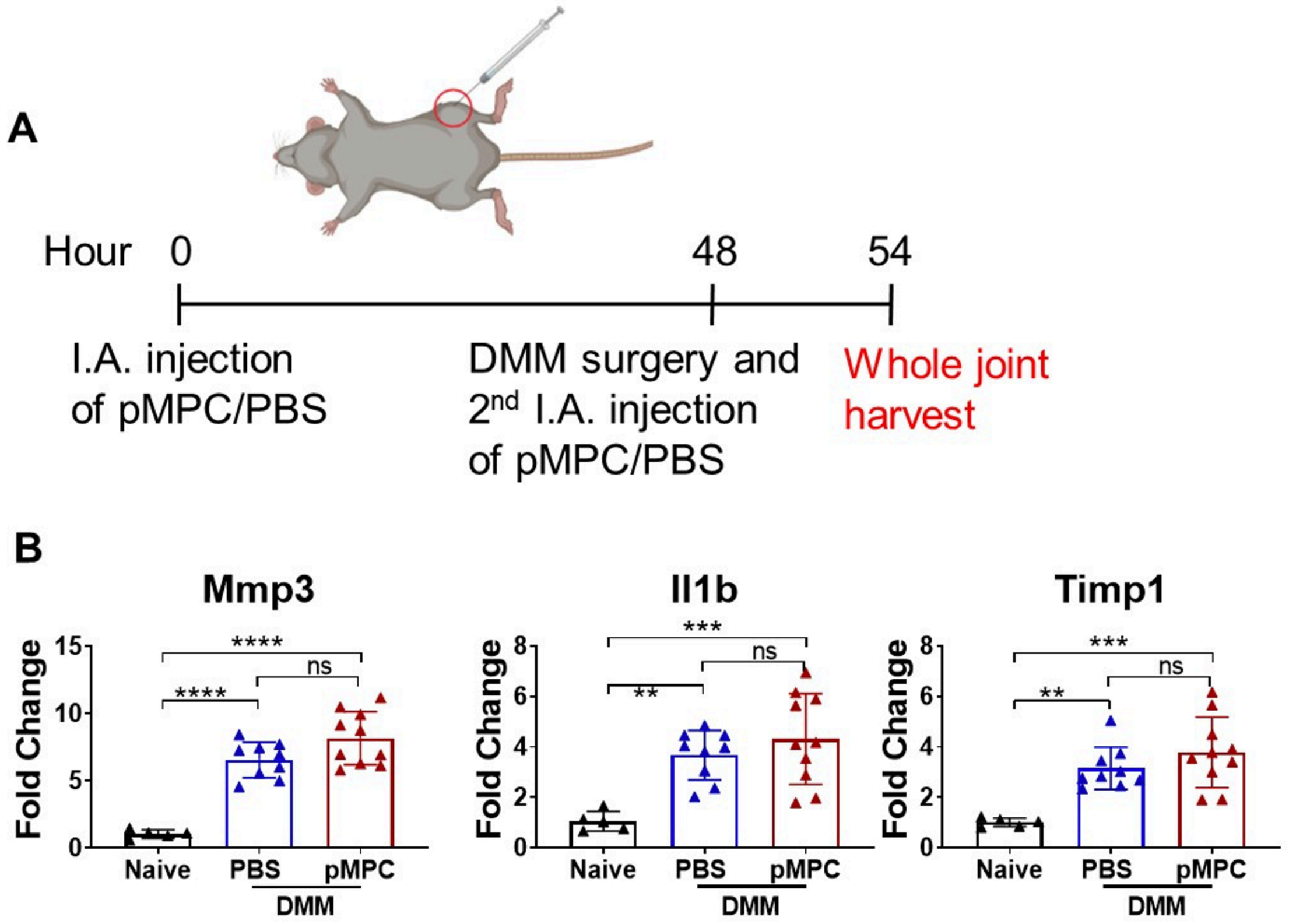

**Fig. S3.** pMPC liposomes do not regulate the expression of the mechano-responsive genes in whole knee joints post-DMM surgery. (A) Schematic of the time course for the experiments. 30 mM of pMPC or PBS vehicle control was injected into the right knee joints of 11-week old male mice through intra-articular injection. Naïve mouse joints were injected with PBS control. 48 hours later, DMM surgery and another intra-articular injection were performed on the same joints. 6 hours post-surgery, whole joints were collected for RNA extraction. (B) Quantification of mRNA expression for mechano-responsive genes *Mmp3, Il1b and Timp1*. Error bars denote mean ± SEM. Results were expressed relative to *18s*, and p values were indicated for each gene using one-way ANOVA with Bonferroni's post hoc analysis. **$p < 0.01$, ***$p < 0.001$, and ****$p < 0.0001$. The increase of all three genes was observed following DMM with PBS (fig. S3B) (fold change *Mmp3*, 6.55 ± 2.19; *Il1b*, 3.68±1.01, *Timp1*, 3.16 ± 0.88), normalised to the naïve (unoperated) control (*Mmp3*,1.04 ± 0.53; *Il1b,* 1.05 ± 0.56; *Timp1*, 1.01 ± 0.50). No difference however was observed in the expression level of genes between PBS and pMPC (*Mmp3,* 8.16 ± 2.99; *Il1b,* 4.32 ± 1.86 *Timp1*, 3.79 ± 1.44) treated groups (fig. S3B)

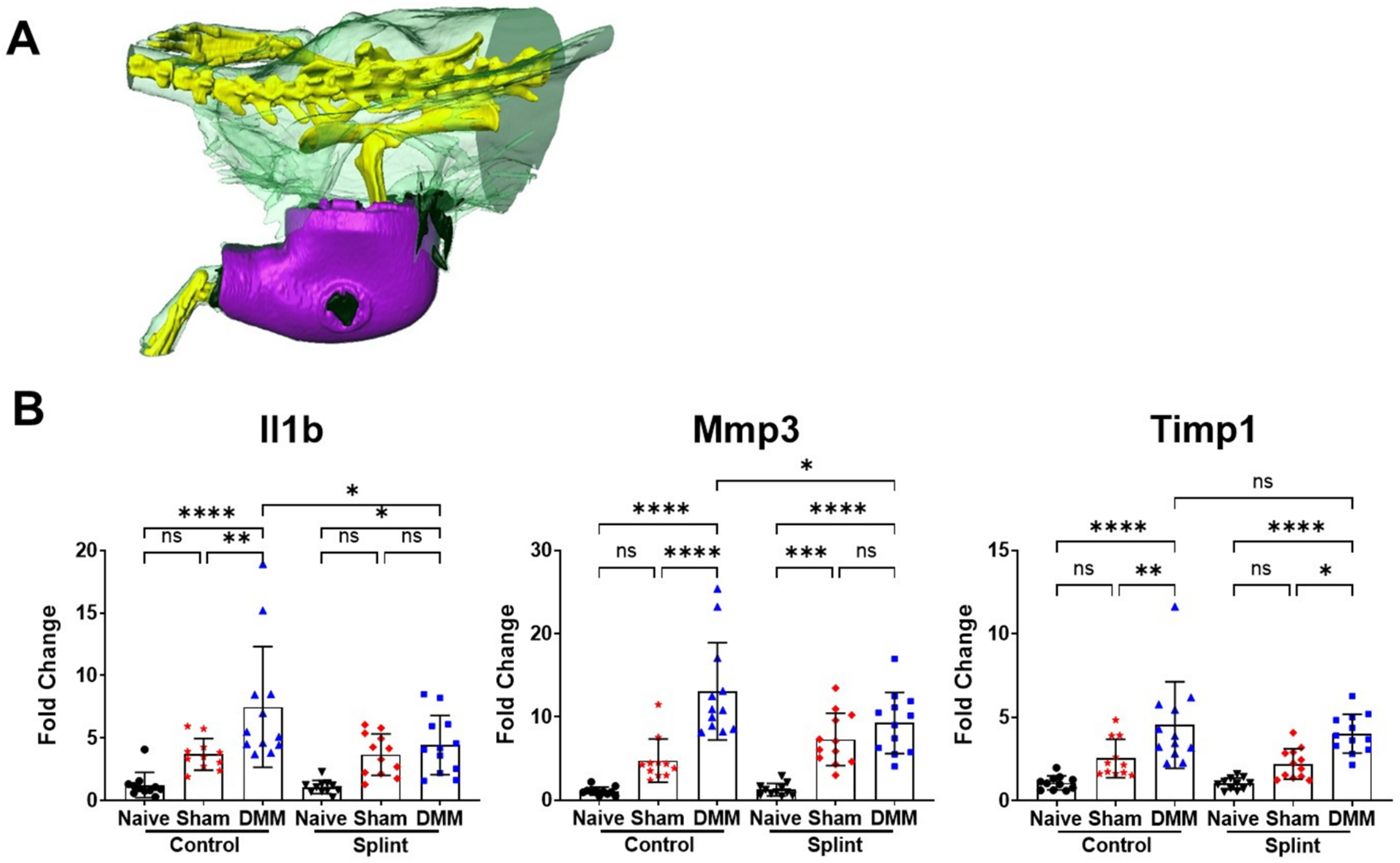

Fig. S4. Suppression of shear-stress induced genes after DMM when mouse joints were splinted. The right knee joints of male C57BL/6 mice under went sham or DMM surgery followed by immediate application of the splint. 6-hour post-surgery, whole joints were collected for RNA extraction. (A) Micro CT scan of hindquarters of a mouse showing the splint (purple), mouse soft tissues (light green) and skeleton (yellow). Each splint was scaled to hold the knee joint in a fixed position that allowed weight bearing (joint compression) but prevented knee flexion. An injection port was designed on the lateral surface of the splint to allow for safe injection of silicone (black) between the splint and mouse soft tissue, which ensures immobilisation of knee joint and any relative splint movement. B. Expression of shear-responsive genes *Mmp3* and *Il1b*, and non-shear responsive gene *Timp1* in the whole joint 6h post DMM surgery with or without splint. Error bars denote mean ± SD. Results were expressed relative to *18s*, and p values were indicated for each gene using one-way ANOVA with Bonferroni's post hoc analysis. n= 12, *p < 0.05, **p < 0.01, ***p < 0.001 and ****p < 0.0001. The splint immobilization thus suppressed shear associated genes *MMp3* (fold change 13.08 ± 5.84) and *Il1b* (7.48 ± 4.83) down to *Mmp3*, 9.27±3.67 and *Il1b,* 4.43 ± 2.36, but had no effect on compression activated gene, *Timp1 (*DMM, 4.54 ± 2.59 to Splint-DMM, 4.01 ± 1.17 (fig S4B).